\documentclass[aps,preprint,preprintnumbers,nofootinbib,showpacs,superscriptaddress]{revtex4-1}
\usepackage{graphicx,color,amsmath}

\begin{document}
\title{Teleparallel Dark Energy with Purely Non-minimal Coupling to Gravity}
\author{Je-An Gu}
\email{jagu@ntu.edu.tw} %
\affiliation{Leung Center for Cosmology and Particle
Astrophysics, National Taiwan University, Taipei, 10617 Taiwan(R.O.C)}
\author{Chung-Chi Lee}
\email{g9522545@oz.nthu.edu.tw}
\affiliation{Department of Physics, National Tsing Hua University, Hsinchu, Taiwan 300}
\author{Chao-Qiang Geng}
\email{geng@phys.nthu.edu.tw}
\affiliation{Department of Physics, National Tsing Hua University, Hsinchu, Taiwan 300}
\affiliation{National Center for Theoretical Sciences, Hsinchu, Taiwan 300}
\begin{abstract}
We propose the simplest model of teleparallel dark energy with
purely a non-minimal coupling to gravity but no self-potential,
a single model possessing various interesting features:
simplicity, self-potential-free, the guaranteed late-time
cosmic acceleration driven by the non-minimal coupling to
gravity, tracker behavior of the dark energy equation of state
at earlier times, a crossing of the phantom divide at a late
time, and the existence of a finite-time future singularity.
We find the analytic solutions of the dark-energy scalar field
respectively in the radiation, matter, and dark energy
dominated eras, thereby revealing the above features. We
further illustrate possible cosmic evolution patterns and
present the observational constraint of this model obtained by
numerical analysis and data fitting.
\end{abstract}

\pacs{95.36.+x, 04.50.Kd, 98.80.Es}

\maketitle

\section{Introduction} \label{sec:introduction}

The accelerating expansion of the present universe discovered
in 1998 is one of the most important puzzles yet to be
solved~\cite{astro-ph/9805201}. Its solution hopefully will
lead us to a new revolution in physics of this century. A
positive cosmological constant, as a geometrical and/or an
energy component of the universe, gives the simplest
explanation that fits the current observational results. In
addition, an energy source of anti-gravity dubbed ``dark
energy'' and the modification of gravity provide two intriguing
approaches to the solution~\cite{Copeland:2006wr}.

A simple realization of dynamical dark energy is given by a
scalar field minimally coupled to gravity, where the cosmic
acceleration is driven by potential
energy~\cite{scalar1,scalar2} or noncanonical kinetic
energy~\cite{ke} of the scalar field. For an alternative
explanation from modified gravity one may introduce new degrees
of freedom, modify the gravity action, or even change the
formalism. A simple extension of general relativity (GR) in the
first scenario is to introduce a scalar field non-minimally
coupled to gravity, e.g., the scalar-tensor
theory~\cite{non-f(R)}. Models with non-minimal derivative
couplings have also been proposed~\cite{non-d}. For the second
scenario a straightforward modification is to invoke a
nonlinear function of the Ricci scalar in the gravity action to
replace the linear function in GR, i.e., the $f(R)$
theory~\cite{f(R)} that can be regarded as one of the
scalar-tensor theories. As to the change of the formalism, an
example is teleparallel gravity that is formulated with torsion
and a curvatureless connection.

Teleparallel gravity can be equivalent to GR when the gravity
action invokes simply a linear function of the torsion scalar.
To make an extension of teleparallel gravity as modified
gravity, i.e.\ inequivalent to GR, one may follow the two
scenarios raised above. A simple extension can be made by
introducing a scalar field non-minimally coupled to
teleparallel gravity. This can be regarded as a
``scalar-teleparallel theory'' of gravity, a modification of
teleparallel gravity analogous to the scalar-tensor theory as a
modification of GR. It has recently been proposed as an
alternative dark energy
model~\cite{Geng:2011aj,Geng:2011ka,Wei:2011yr,Xu:2012jf} and
dubbed ``teleparallel dark energy.'' For the other scenario a
straightforward modification of the gravity action can be made
by invoking a nonlinear function of the torsion scalar, i.e.,
the $f(T)$ theory (for a review, see \cite{Bamba:2012cp}).

In this paper we focus on teleparallel dark energy. The
distinct behavior of dark energy in this model has been
investigated in \cite{Geng:2011aj,Geng:2011ka} where several
different potential terms for the self-interaction of the
scalar field are chosen. For lack of a guiding principle of
choosing the self-potential, in this paper we propose the
simplest model with no potential. That is, we investigate the
``minimal'' model of teleparallel dark energy where the scalar
field is canonical, massless and noninteracting but
non-minimally coupled to teleparallel gravity. With this simple
model we attempt to manifest the effect of the non-minimal
coupling on the cosmic acceleration, in contrast to the
conventional models that attribute the acceleration to
potential energy or noncanonical kinetic energy.

In this model we find the analytic solutions of the scalar
field in the radiation, matter, and dark energy dominated eras,
respectively. These solutions largely facilitate the
theoretical studies on the dark energy behavior and help to
reveal various interesting features of this model. In addition
to the analytic solutions, we also numerically analyze this
model and illustrate possible evolution patterns. As to be
shown, the numerical solutions truly manifest the features read
from the analytic solutions.
We then present the observational constraint of this model
obtained by the data fitting with the type Ia supernova
(SNIa)~\cite{Perivolaropoulos:2004yr}, baryon acoustic
oscillation (BAO)~\cite{astro-ph/0501171,Eisenstein:1997ik},
and cosmic microwave background (CMB)~\cite{Bond:1997wr,
Hu:1995en} observational results.

\section{Teleparallel Dark Energy without Self-Potential} \label{sec:TeleDE}

Teleparallel dark energy is played by a canonical scalar field
non-minimally coupled to gravity in the framework of
teleparallel gravity, which  is formulated with the veirbein
field $\mathbf{e}_A (x^{\mu})$, the metric deduced from
veirbein $g_{\mu \nu}(x) = \eta_{AB} e^A_{\mu} e^B_{\nu}$, the
curvatureless Weitzenb\"{o}ck connection
$\overset{\mathbf{w}}{\Gamma}{}^\lambda_{\nu\mu} \equiv
e^{\lambda}_A \partial_{\mu} e^A_{\nu}$, and the corresponding
torsion tensor
\begin{equation}  \label{eq:torsion2}
{T}^\lambda_{\:\mu\nu}\equiv \overset{\mathbf{w}}{\Gamma}{}^\lambda_{
\nu\mu}-\overset{\mathbf{w}}{\Gamma}{}^\lambda_{\mu\nu}
=e^\lambda_A\:(\partial_\mu
e^A_\nu-\partial_\nu e^A_\mu)\,.
\end{equation}
The action of teleparallel dark energy reads
\begin{equation} \label{eq:action-TDE}
S=\int d^{4}x \, e \left[ \frac{T}{2\kappa^{2}}
+ \frac{1}{2} \left(\partial_{\mu}\phi\partial^{\mu}\phi+\xi
T\phi^{2}\right) - V(\phi) +\mathcal{L}_m \right] ,
\end{equation}
where $e\equiv \text{det}(e_{\mu}^A) = \sqrt{-g}$, $T$ is the
torsion scalar defined by
\begin{eqnarray}
T &\equiv& \frac{1}{4}T^{\rho \mu \nu }T_{\rho \mu \nu }
+ \frac{1}{2}T^{\rho \mu \nu}T_{\nu \mu \rho }
- T_{\rho \mu }^{\ \ \rho }T_{\ \ \ \nu }^{\nu \mu} \, ,
\end{eqnarray}
$\phi$ is the scalar field, $V(\phi)$ the self-potential,
$\mathcal{L}_m$ the matter Lagrangian, and $\xi$ the
non-minimal coupling constant. In this paper we investigate the
simplest model free of self-potential, i.e., hereafter
$V(\phi)=0$.

For a flat, homogeneous and isotropic background where the
space-time is described by the flat Robertson-Walker metric
$ds^2 = dt^2 - a^2(t) \delta_{ij} dx^i dx^j$ and the scalar
field is simply time-dependent, the scalar field and the
gravitational field equations \cite{Geng:2011aj} read
\begin{eqnarray} \label{eq:eom}
&& \ddot{\phi} + 3H\dot{\phi} + 6\xi H^2\phi = 0 \, ,
\nonumber \\
&& H^2 \equiv \left(\frac{\dot{a}}{a}\right)^2
= \frac{\kappa^2}{3} \left( \rho_\phi
+ \rho_\textrm{m} + \rho_\textrm{r} \right) ,
\nonumber \\
&& \dot{H} = -\frac{\kappa^2}{2} \left( \rho_{\phi} + p_{\phi}
+ \rho_\textrm{m} + 4\rho_\textrm{r}/3 \right) ,
\end{eqnarray}
where the matter energy density $\rho_\textrm{m} \propto
a^{-3}$, the radiation energy density $\rho_\textrm{r} \propto
a^{-4}$, and the energy density and pressure of the scalar
field are given by
\begin{eqnarray}
\rho_{\phi} &=& \frac{1}{2} \dot{\phi}^2  - 3\xi H^2 \phi^2 \, ,
\nonumber \\
p_{\phi} &=& \frac{1}{2} \dot{\phi}^2 + 3\xi H^2 \phi^2 + 2\xi \frac{d}{dt}(H\phi^2) \, .
\end{eqnarray}
We note that for $\xi < 0$, the non-minimal coupling may result in
 negative pressure, thereby anti-gravity, and meanwhile a
positive energy density. In particular, the contribution from
the term $3\xi H^2 \phi^2$ entails the cosmological-constant
equation of state $w=-1$ and the other term $2\xi
d(H\phi^2)/dt$ provides no energy density but negative pressure
when $H\phi^2$ increases with time. Henceforth we consider the
case where $\xi < 0$ in order to guarantee the positiveness of
the energy density. We set the present scale factor $a_0 = 1$
for convenience and without loss of generality.

\section{Analytic Solutions} \label{sec:solution}

Here we present the analytic solutions of the scalar field in
two cases: (a) the case where $H \propto 1/t$, including the
radiation-dominated (RD) and matter-dominated (MD) eras, and
(b) the scalar-field-dominated (SD) era.

\vspace{0.5em}%
\noindent \textbf{(a) $H = \alpha/t$, i.e.\ $a(t) \propto
t^{\alpha}$, with constant $\alpha$:}
\begin{equation}
\label{eq:phit}
\phi (t) = C_1 t^{l_1} + C_2 t^{l_2} ,
\end{equation}
where $C_{1,2}$ are constants and
\begin{eqnarray}
l_{1,2} &=&  \hspace{0.7em}
         \frac{1}{2} \left[ \pm \sqrt{(3\alpha-1)^2 - 24\xi \alpha^2} - (3\alpha-1) \right].
\label{eq:l12}
\end{eqnarray}
In RD, $\alpha=1/2$; in MD, $\alpha=2/3$. For $\xi<0$, the
power-index $l_1$ is positive and $l_2$ negative, corresponding
to an increasing mode and a decreasing mode, respectively.

Hereafter we take into consideration only the increasing mode,
i.e., $\phi (t) = C_1 t^{l_1}$. The energy density,  pressure,
and equation of state corresponding to this solution are given
by
\begin{eqnarray} \label{eq:RMDsoln}
\rho_{\phi} &=& \left[ l_1^2 /2 - 3\xi \alpha^2 \right] C_1^2 t^{2(l_1-1)} ,
\nonumber\\
p_{\phi}    &=& \left[ l_1^2 /2 + 3\xi \alpha^2 + 2\xi \alpha(2l_1-1) \right] C_1^2 t^{2(l_1-1)} ,
\nonumber \\
w_{\phi} &\equiv& \frac{p_{\phi}}{\rho_{\phi}} = 1+\frac{4\xi \alpha}{l_1} = -1+\frac{2(1-l_1)}{3\alpha} .
\end{eqnarray}
Consequently, one obtains
\begin{eqnarray}
\textrm{RD:} &\;\;&
w_{\phi} = {1\over 3}\left( 2 - \sqrt{1-24\xi}\right), \;\,
\rho_{\phi} \propto a^{-5+\sqrt{1-24\xi}} \, ; \\
\textrm{MD:} &\;\;&
w_{\phi} = {1\over 2}\left( 1- \sqrt{1-32\xi/3} \right), \;\,
\rho_{\phi} \propto a^{(-9 +\sqrt{9-96\xi} \, )/2} \, .
\end{eqnarray}

To sketch the behavior of teleparallel dark energy in RD and
MD, in Table~\ref{table:models} we consider four different
$\xi$ as the examples for demonstration. The $\xi$ values under
consideration are within the range $[-1,0)$ because the case
where $\xi<-1$ is rather disfavored by observational results
about the cosmic expansion. The four values of $\xi$ are so
chosen that in RD or MD the dark energy density $\rho_{\phi}$
behaves like the familiar energy sources such as radiation,
matter, or a cosmological constant: %
(i) When $\xi=0^{-}$, teleparallel dark energy tracks the
dominant energy components respectively in RD and MD, a
behavior similar to the tracker quintessence
\cite{TrackerQ} 
(ii) When $\xi=-1/8$, in RD $\rho_{\phi}$ behaves like matter
and in MD it drops slower than matter. %
(iii) When $\xi=-3/4$, in RD $\rho_{\phi}$ decreases rather
slowly and in MD it is constant. %
(iv) When $\xi=-1$, in RD $\rho_{\phi}$ is constant and in MD
it increases as phantom.
\begin{table}[htbp]
\caption{Explicit solutions of the  scalar field in RD and MD
with different values of $\xi$.} \vskip 0.2in
\label{table:models}
\begin{tabular}{|c||c|c||c|c||c|c||c|c|} \hline
$\xi$ & \multicolumn{2}{c||}{(i) $0^{-}$} & \multicolumn{2}{c||}{(ii) $-1/8$}
& \multicolumn{2}{c||}{(iii) $-3/4$} & \multicolumn{2}{c|}{(iv) $-1$} \vline
\\ \hline \hline
 Era & RD&MD & RD&MD & RD&MD & RD&MD
\\ \hline
$l_1$& $ -3\xi$ & $-8\xi/3$ & $+1/4$ & $+0.264$ & $0.840$ & $+1$ & $+1$ & $+1.208$
\\ \hline
$w_{\phi}$& $1/3+4\xi$ & $-8\xi/3$ & $0$ & $-0.264$ & $-0.786$ & $-1$ & $-1$ & $-1.208$
\\ \hline
$\rho_{\phi} \propto$& $a^{-4}a^{-12\xi}$ & $a^{-3}a^{-8\xi}$ & $a^{-3}$ & $a^{-2.21}$ & $a^{-0.641}$ & Constant & Constant &$a^{0.623}$
\\ \hline
\end{tabular}
\label{Table}
\end{table}

The analytic solution, as well as the examples in
Table~\ref{table:models}, shows several interesting features:
The dark energy equation of state $w_{\phi}$ is a constant
simply determined by the sole model parameter $\xi$ but
insensitive to the initial condition of the scalar field, i.e.,
a tracker behavior of the dark energy equation of
state.\footnote{This is different from tracker quintessence
\cite{TrackerQ} 
where it is the scalar field (thereby the dark energy density
and its equation of state) that possesses the tracker
behavior.} For $\xi<0$, $w_{\phi}$ is smaller than that of the
dominant energy component, $w_{\alpha}=-1+2/(3\alpha)$, at all
times in RD and MD. Accordingly, the dark energy density
$\rho_{\phi}$ decreases slower than the dominant energy density
$\rho_{\alpha}$ and therefore the late-time domination of dark
energy (SD) is guaranteed. For larger $|\xi|$, $w_{\phi}$ is
smaller and $\rho_\phi / \rho_\alpha$ increases faster, and
therefore SD comes sooner for a fixed initial condition.

The analytic solutions in RD and MD describe the universe at
earlier times. At later times, the universe is 
dominated by both matter and dark energy (MSD), e.g., in the
present epoch, and then dominated solely by dark energy (SD) in
the future. In the following we attempt to analytically
integrate the field equations 
in both of the later eras (MSD and SD).


\vspace{0.5em} %
\noindent \textbf{(b) Scalar field domination:} %

For the universe dominated by matter and dark energy, we deduce
an integrable equation from Eq.\ (\ref{eq:eom}):
\begin{eqnarray}
\frac{d}{dt}\left[ F(\phi) a^3H \right]
&=& \frac{\kappa^2}{2}\rho_\textrm{m} a^3 = \frac{3}{2}H_0^2 \Omega_\textrm{m0}\,,
\nonumber\\
F(\phi) &\equiv & 1+\kappa^2\xi\phi^2 ,
\end{eqnarray}
where $H_0$ and $\Omega_\textrm{m0}$ are  the Hubble constant
and the matter density fraction at the present time,
respectively. Its solution gives a relation:
\begin{equation} \label{eq:relation-in-MSD}
F(\phi)a^3H = \frac{3}{2}H_0^2 \Omega_\textrm{m0} t + C_3 \, ,
\end{equation}
where $C_3$ is an integration constant. This can hardly be
further integrated analytically.

For the universe dominated by dark energy solely, Eq.\
(\ref{eq:relation-in-MSD}) gives a simpler relation with
$\Omega_\textrm{m0}=0$:
\begin{equation}
F(\phi)a^3H = C_3 \, .
\end{equation}
  From this relation and the field equations we obtain the second
integrable equation,
\begin{equation}
\frac{1}{F(\phi)}\left(\frac{d\phi}{d\ln a}\right)^2
= \frac{6}{\kappa^2} \, ,
\end{equation}
from which one can straightforwardly obtain the solution,
\begin{eqnarray}
\phi (a) & =& \pm \frac{\sin \theta}{\sqrt{-\kappa^2\xi}} \, ,
\nonumber\\
\theta (a) &\equiv &\sqrt{-6\xi} \ln a + C_4 \, ,
\end{eqnarray}
and then the equation of state,
\begin{equation}
\label{eq:wdiv}
w_{\phi} = -1 - \sqrt{-32\xi /3} \tan \theta \,,
\end{equation}
where $C_4$ is an integration constant.

This solution predicts a future singularity where both the
expansion rate $H$ and the dark energy equation of state
$w_\phi$ go to infinity. It is one kind of finite-time future
singularities, i.e.\ ``type III'' singularity classified in
Ref.\ \cite{Nojiri:2005sx}. The singularity occurs when $\theta
= (n+1/2) \pi$ for integer $n$, i.e., $\phi = \pm
1/\sqrt{-\kappa^2 \xi}$ and $F(\phi)=0$. Between two
singularities, e.g., for $-\pi /2 < \theta < \pi /2$, $\phi$ is
monotonic and $w_{\phi}$ monotonically decreases from $+\infty$
to $-\infty$ along with the expansion, experiencing the
crossing of the phantom divide $w_{\phi}=-1$ at $\theta = 0$.

\section{Possible Evolution Patterns and Data Fitting} \label{sec:expansion-patterns}

Here we illustrate the evolution patterns of the dark energy
equation of state $w_{\phi}$ for a wide range of initial
conditions by numerically solving the field equations
(\ref{eq:eom}). These numerical solutions will manifest the
interesting features we have read from the analytic solutions,
particularly the feature that $w_{\phi}$ approaches the
constant tracker values 
in RD and MD, leaves the tracker and starts decreasing when the
universe is leaving MD for SD, keeps decreasing and crosses the
phantom divide in SD, and goes to negative infinity at a finite
time eventually. In addition, we confront this
self-potential-free teleparallel dark energy model with the
observational data that are relevant to the cosmic expansion,
including SNIa \cite{Perivolaropoulos:2004yr}, BAO
\cite{astro-ph/0501171,Eisenstein:1997ik} and CMB
\cite{Bond:1997wr, Hu:1995en},
thereby obtaining the constraint on the sole model parameter,
i.e.\ the non-minimal coupling constant $\xi$, as well as the
cosmological parameters.

\begin{figure}[h!]
\includegraphics[width=7cm]{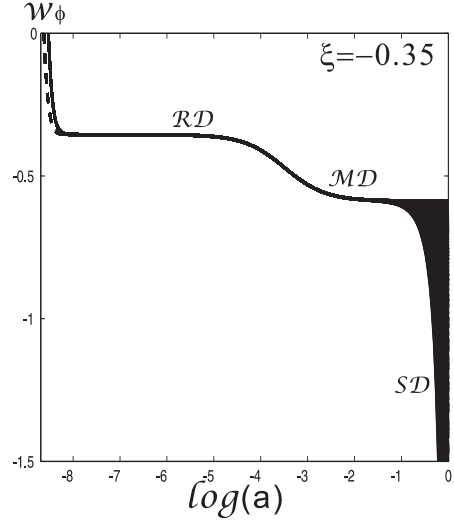}
\caption{\label{fig:w-xi04}
Evolution of the dark energy equation of state,
$w_{\phi}(\log(a))$, with $\xi = -0.35$. The black area in SD
is formed by the trajectories of $w_{\phi}(\log(a))$ w.r.t.\ a
wide range of initial conditions.}
\end{figure}

\begin{figure}[h!]
\includegraphics[width=7cm]{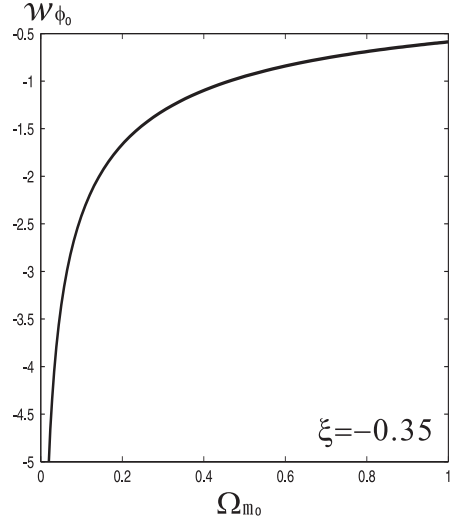}
\caption{\label{fig:Om&w-xi04}
The present values of $(\Omega_m , w_{\phi})$ for a wide range
of initial conditions with $\xi = -0.35$. The curve shows a
relation between these two quantities, as a consequence of the
tracker behavior of $w_{\phi}$ for a flat universe with given
$\xi$.}
\end{figure}

For illustrating the evolution patterns of $w_{\phi}$
numerically, 
we consider the case where $\xi = -0.35$ and show
$w_{\phi}(\log(a))$ in Fig.~\ref{fig:w-xi04} for different
initial conditions. For the same case we also show the present
values of $(\Omega_m,w_{\phi})$ in Fig.~\ref{fig:Om&w-xi04}. We
consider two sets of initial conditions at the initial time
$\log(a) \simeq -8.69$ (i.e.\ $\ln (a) = -20$): (i) $\kappa
\phi =0$ and $\kappa \phi^{\prime} \equiv d (\kappa \phi) /
d\ln a \in [10^{-50}, 2.5 \times 10^{-8}]$ (solid line), and
(ii) $\kappa \phi^{\prime} = 0$ and $\kappa \phi \in [
10^{-50}, 1.2 \times 10^{-8} ]$ (dashed line). For these two
kinds of initial conditions, initially $w_{\phi}$ is unity and
$1/3$, respectively. Then, as shown in Fig.~\ref{fig:w-xi04},
it quickly evolves to the RD tracker value, $-0.36$, along with
the domination of the increasing mode of $\phi(t)$-field over
the decreasing mode. It reaches the tracker value around
$\log(a)=-8$. When the universe gradually leaves RD and enters
MD, 
$w_{\phi}$ gradually decreases to the MD tracker value,
$-0.58$. Later, when teleparallel dark energy starts to
dominate, $w_{\phi}$ leaves the MD tracker and decreases
monotonically. Finally, $w_{\phi}$ crosses the phantom divide
($w_{\phi}=-1$) and goes to the singularity as expected.

Because of the tracker behavior of $w_{\phi}$, the evolution of
dark energy is simply determined by two parameters, the model
parameter $\xi$ and the integration constant $C_1$ appearing in
Eq.~(\ref{eq:phit}), where $C_1$ may be regarded as the only
important initial-condition parameter. Accordingly, for a flat
universe with given $\xi$, the present value of $w_{\phi}$ and
the present matter density fraction $\Omega_{m0}$ are related
by a single parameter $C_1$. This relation is shown in Fig.\
\ref{fig:Om&w-xi04}, where both $\Omega_{m0}$ and $w_{\phi 0}$
change with the initial condition, i.e.\ the value of $C_1$.
The present value of $w_\phi$ has an upper bound that depends
on $\xi$ and is set by the tracker value of $w_\phi$ in MD.
Note that the two different sets of initial conditions give the
same relation curve.

\begin{figure}[h!]
\centering
\includegraphics[width=2.5in]{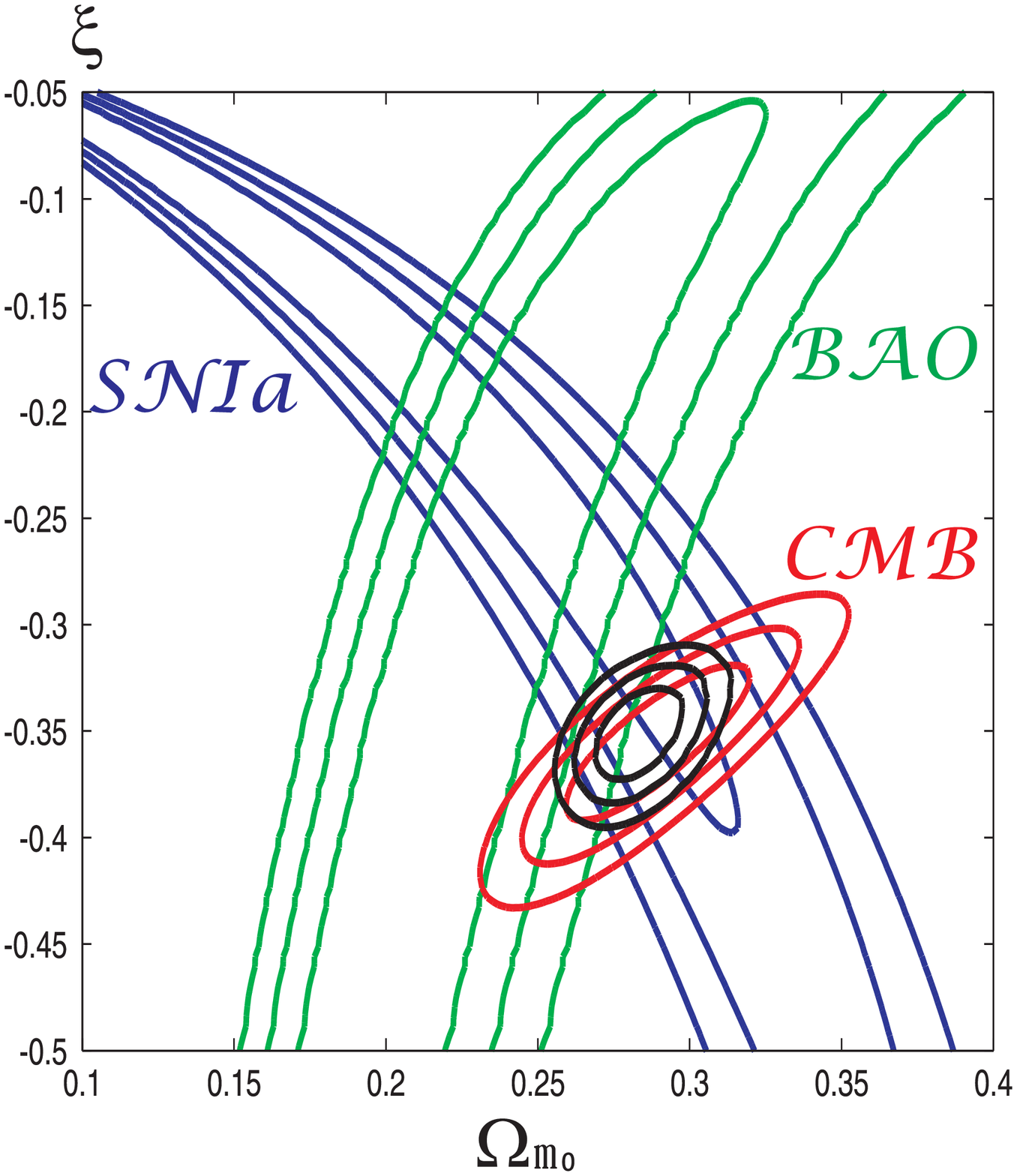}
\includegraphics[width=2.5in]{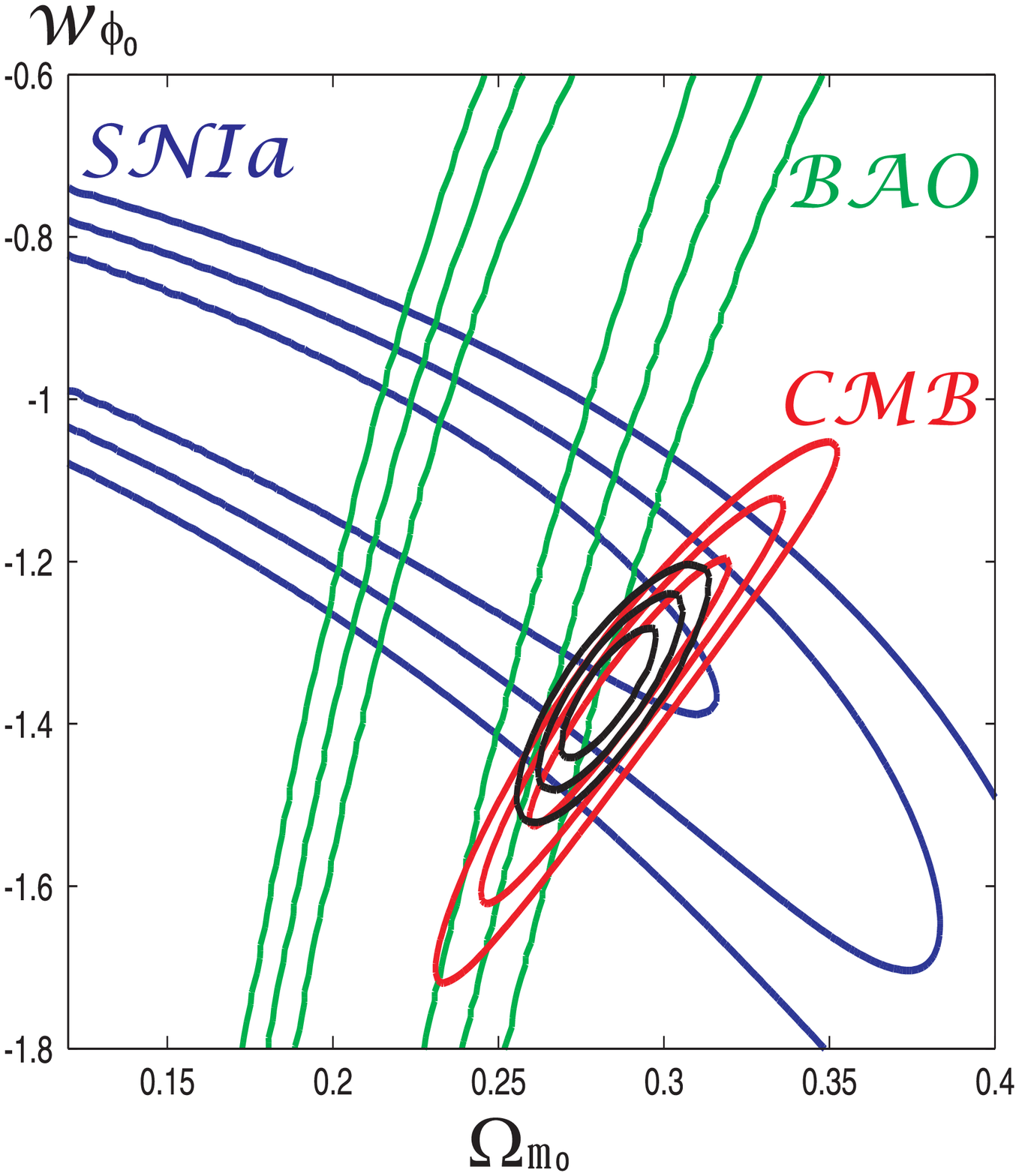}
\caption{
The $1\sigma$--$3\sigma$ confidence regions in  $(\Omega_{m0},\xi)$
(left panel) and $(\Omega_{m0},w_{\phi 0})$ (right panel) obtained
from the SNIa (blue), BAO (green), CMB (red), and the combined
(black) data.} \label{fig:fitting}
\end{figure}

Following the same procedure of data fitting in
Ref.~\cite{Geng:2011ka}, we obtain the observational constraint
of our model.
Figure \ref{fig:fitting} shows the $1\sigma$--$3\sigma$
confidence regions in $(\Omega_{m0},\xi)$ (left panel) and
$(\Omega_{m0},w_{\phi 0})$ (right panel) obtained from the
SNIa~(blue), BAO~(green), CMB~(red), and the combined~(black)
data. The best fit to the combined data locates at $\xi \simeq
-0.35$ and $\Omega_{\textrm{m0}} \simeq 0.28$. The three data
sets do not give a good concordance region, indicating the
imperfection in simultaneously fitting these data sets. This is
possibly due to the feature of this model that $\rho_{\phi}$
may not be negligible at early times and accordingly may
significantly affect the CMB result. We also remark that in
contrast the data can be fitted well in the teleparallel dark
energy models with potentials in Refs.\
\cite{Geng:2011aj,Geng:2011ka}. This is expected because the
models with potentials are more flexible than our simple
potential-free model when fitting data.



\section{Summary} \label{sec:summary}

We propose the simplest model of teleparallel dark energy where
dark energy is played by a scalar field that is canonical,
massless and noninteracting but purely non-minimally coupled to
teleparallel gravity. We have found the analytic solutions of
the scalar field, thereby obtaining the analytic behaviors of
the dark energy equation of state, in the radiation, matter,
and dark energy dominated eras, respectively. With the analytic
solutions we have shown various interesting features of this
simple model:
\begin{itemize}
\item The cosmic acceleration is driven by the non-minimal
    coupling to gravity, but not by potential energy or
    noncanonical kinetic energy as invoked in the
    conventional scalar-field dark energy models.
\item The domination of dark energy, the crossing of the
    phantom divide, and therefore the occurrence of the
    cosmic acceleration at late times are destined.
\item In the radiation and matter dominated eras the dark
    energy equation of state is roughly a constant and has
    tracker behavior, i.e.\ being insensitive to the
    initial condition. 
\item In the dark energy dominated era the dark energy
    equation of state decreases monotonically and crosses
    the phantom divide at a late time; its value strongly
    depends on the initial condition of the scalar field.
\item The universe will meet a future singularity with the
    expansion rate going to infinity within a finite time.
\item This model has simply one free parameter, the
    non-minimal coupling constant $\xi$, for a spatially
    flat FLRW universe with the present matter density
    $\Omega_\textrm{m0}$ fixed.
\end{itemize}

We have fitted the model parameter $\xi$ with the observational
data of SNIa, BAO and CMB in cosmology. The concordance region
for all these three data sets is only at the $3 \sigma $ level.
It may indicate a slight incapability for this model to
describe the early universe. This requires further examination.
For further investigations, in addition to cosmological
observations, the local gravity tests are expected to give
significant constraints on teleparallel dark energy and
therefore should be worth detailed studies.

\begin{acknowledgments}
The work was supported in part by
National Center of Theoretical Science and by National Science
Council of R.O.C.\ under Grants Nos.\ NSC98-2112-M-002-007-MY3
(JAG) and NSC-98-2112-M-007-008-MY3 (CCL and CQG).
\end{acknowledgments}

%

\end{document}